\documentclass[11pt,onecolumn,amssymb,nofootinbib]{revtex4}
\usepackage{amsmath, amsthm, amscd, amssymb}
\usepackage{bm}
\usepackage{bbm}

\begin{document}

\title{\bf Incorporating gravity into trace dynamics: the induced gravitational action}

\author{Stephen L. Adler}
\email{adler@ias.edu} \affiliation{Institute for Advanced Study,
Einstein Drive, Princeton, NJ 08540, USA.}

\begin{abstract}
We study the incorporation of gravity into the trace dynamics
framework for classical matrix-valued fields, from which we have proposed
that quantum field theory is the emergent thermodynamics, with state vector reduction
arising from fluctuation corrections to this thermodynamics.
We show that the metric must be incorporated
as a classical, not a matrix-valued, field, with the source for gravity
the exactly covariantly conserved trace stress-energy tensor of the matter fields.
We then study corrections
to the  classical gravitational action induced by the dynamics of the
matrix-valued matter fields, by examining the average over the trace dynamics
canonical ensemble of the matter field action, in the
presence of a general background metric.  Using constraints from global Weyl scaling
and three-space general coordinate transformations, we show that to zeroth order in
derivatives of the metric, the induced gravitational action in the preferred rest
frame of the trace dynamics canonical ensemble must  have the form
\begin{equation}\label{eq:action}
\Delta S=\int d^4x (^{(4)}g)^{1/2}(g_{00})^{-2} A\big(g_{0i}g_{0j}g^{ij}/g_{00},D^ig_{ij}D^j/g_{00},g_{0i}D^i/g_{00}\big)~~~,
\end{equation}
with $D^i$ defined through the co-factor expansion of $^{(4)}g$ by $^{(4)}g/{^{(3)}g}=g_{00}+g_{0i}D^i$,
and with $A(x,y,z)$ a general function of its three arguments.
This action has ``chameleon-like'' properties:  For the Robertson-Walker cosmological
metric, it {\it exactly} reduces to a cosmological constant, but for the Schwarzschild
metric it diverges as $(1-2M/r)^{-2}$ near the Schwarzschild radius, indicating that
it may substantially affect the horizon structure.
\end{abstract}

\maketitle

\section{Introduction}

In papers culminating in a book   \cite{adlermillard}--\cite{adlerbook}, we proposed ``trace dynamics'' as the fundamental
pre-quantum dynamics of matter degrees of freedom.  In this dynamics, the
matter fields are non-commuting matrix-valued fields, with cyclic permutation
under a trace action resolving factor-ordering problems.  We identified globally conserved
quantities and used them to construct a canonical ensemble for the statistical dynamics
of trace dynamics.  We then argued, with approximations that remain to be justified, that
the statistical thermodynamics of trace dynamics gives rise to quantum field theory, with
fluctuation corrections to this thermodynamics leading to state vector reduction in
measurements.

We did not, however, address the issue of incorporating gravity into trace dynamics, and
that is the purpose of this paper.  We shall see that without knowing the precise underlying
trace dynamics action, restrictive qualitative statements can be made. The organization of
this paper is as follows.  In Sec. 2 we give a very brief survey of trace dynamics.
In Sec. 3 we present arguments indicating that gravity must be incorporated into
trace dynamics as a classical  (a diagonal matrix) field, as opposed to a general matrix-valued field.
We show that this leads to a consistent dynamics for classical gravity coupled to matrix-valued
matter, with the source term for the Einstein equations the exactly covariantly conserved matter trace
stress-energy tensor.
In Sec. 4 we define a matter induced effective action as the canonical ensemble average of the
matter trace action.  In Sec. 5, we  use global Weyl scaling invariance and three-space general coordinate invariance
to derive a general functional form for the structure of the induced effective action
to leading orders in derivatives of the metric.  In Sec. 6 we deduce the rules for using
this frame dependent effective action as a source for the Einstein equations.  In Sec. 7, we show that although this
effective action does not have the structure of a cosmological constant action, on a Robertson-Walker
space-time it {\it exactly} reduces to a cosmological term.  In Sec. 8 we take a first look at the implications
of the effective action for a time-independent, spherically symmetric metric.  In Sec. 9 we make further remarks about the
structure and implications of our results.    In Appendix A we state our notational conventions and give formulas for matter field actions and conserved quantities derived from them. In Appendix B we discuss the construction and properties of the mixed index gravitational pseudotensor that is referenced in the course of our argument.

\section{Brief overview of trace dynamics}

Trace dynamics \cite{adlermillard}--\cite{adlerbook} is a new kinematic framework for pre-quantum dynamics, in which the dynamical variables are pairs
of operator valued variables $\{q_r\}, \,\{p_r\}$, with no assumed {\it a  priori} commutativity properties,
acting on an underlying complex Hilbert space. A theory of dynamical flows can be set up by starting from
an operator Hamiltonian $H[\{q_r\}, \,\{p_r\}]$ and forming the trace Hamiltonian ${\bf H} \equiv {\rm Tr} H$.  Although
noncommutativity of the operator variables prevents one from differentiating $H$ with respect to them, we
can use the cyclic property of the trace to define derivatives of a general trace functional ${\bf A}$  by
forming $\delta {\bf A}$ and cyclically reordering the operator variations $\delta q_r,\, \delta p_r$ to
the right.  This gives the fundamental definition
\begin{equation}\label{eq:def1}
\delta {\bf A} = {\rm Tr} \sum_r \left(\frac{\delta{\bf A}}{\delta q_r}  \delta q_r
+ \frac{\delta{\bf A}}{\delta p_r}  \delta p_r\right) ~~~,
\end{equation}
in which $\delta {\bf A}/\delta q_r$ and $\delta {\bf A}/\delta p_r$ are operators.  Applying this definition
to the trace Hamiltonian, a symplectic dynamics of the operator phase space variables is generated by
the operator Hamilton equations
\begin{equation}\label{eq:hamdynam}
\frac {\delta {\bf H}}{\delta q_r}= -\dot p_r~,~~~\frac {\delta {\bf H}}{\delta p_r}= \epsilon_r\dot q_r~~~,
\end{equation}
with $\epsilon_r = 1 (-1)$ according to whether $q_r,\,p_r$ are bosonic (fermionic).

Substituting Eq. \eqref{eq:hamdynam} into Eq. \eqref{eq:def1}, we see that ${\bf H}$ is a constant of motion
in trace dynamics. Another conserved trace quantity is the trace fermion number {\bf N}.
 An essential feature of trace dynamics is that there are two other conserved quantities.
The first is the traceless anti-self-adjoint operator $\tilde C$ defined by
\begin{equation}\label{eq:cdef}
\tilde C= \sum_{r,B} [q_r,p_r] - \sum_{r,F}\{q_r,p_r\}~~~,
\end{equation}
with the subscripts $B,F$ denoting respectively sums over bosonic and fermionic degrees of freedom.
When the trace Hamiltonian is constructed using only non-operator numerical coefficients, there is
a global unitary invariance for which $\tilde C$ is the conserved Noether charge \cite{adlerkempf}.

A second important
conserved quantity is the natural integration measure $d\mu$ for the underlying operator phase space.
Conservation of $d\mu$ gives a trace dynamics analog of Liouville's theorem, and permits the use of
statistical mechanics methods.  Specifically, the canonical ensemble is given by
\begin{equation}\label{eq:canon}
d\mu \rho=d\mu \rho(\tilde C, \tilde \lambda; {\bf H}, \tau)=\frac {d\mu \exp[-{\rm Tr}( \tilde \lambda \tilde C) -\tau {\bf H}-\eta {\bf N}]}
{\int d\mu \exp[-{\rm Tr}( \tilde \lambda \tilde C) -\tau {\bf H}-\eta {\bf N}]}~~~,
\end{equation}
with the denominator enforcing the normalization condition $\int d\mu \rho=1$.
  The ensemble parameters (generalized
temperatures) are the real numbers $\tau$ and $\eta$, and the anti-self-adjoint operator $\tilde \lambda$, chosen so
that the ensemble averages
\begin{equation}\label{eq:avdef}
\langle {\bf H} \rangle_{\rm AV}= \int d\mu \rho {\bf H}, ~~~\langle \tilde C \rangle_{\rm AV}= \int d\mu \rho \tilde C~~~,\langle {\bf N} \rangle_{\rm AV}= \int d\mu \rho {\bf N}
\end{equation}
have specified values.  Since $\langle \tilde C \rangle_{\rm AV}$ is itself an anti-self-adjoint operator,
it can be brought to the canonical form
\begin{equation}\label{eq:canon0}
\langle \tilde C \rangle_{\rm AV}=i_{\rm eff} D_{\rm eff}~,~~~{\rm Tr}(i_{\rm eff} D_{\rm eff})=0~,~~~~i_{\rm eff}=-i_{\rm eff}^{\dagger}~,~~~i_{\rm eff}^2=-1~,~~~[i_{\rm eff},D_{\rm eff}]=0~~~,
\end{equation}
with $D_{\rm eff}$ a real diagonal and non-negative operator.

The simplest case corresponds to an ensemble that does not favor any state in the underlying Hilbert space over
any other, in which case $D_{\rm eff}$ is a real constant multiple of the unit operator.  This real constant has
the dimensions of action, and plays the role of Planck's constant in the emergent quantum mechanics derived
from the canonical ensemble, so we shall denote it by $\hbar$, giving
\begin{equation}\label{eq:canon1}
\langle{\tilde C} \rangle_{\rm AV}=i_{\rm eff} \hbar~,~~~{\rm Tr}\,i_{\rm eff}=0~~~.
\end{equation}
Since the relations $i_{\rm eff}=-i_{\rm eff}^{\dagger}$ and $i_{\rm eff}^2=-1$ imply that $i_{\rm eff}$ can be
diagonalized to the form $i {\rm diag}(\pm1,\pm1,...,\pm1)$, the condition ${\rm Tr}\,i_{\rm eff}=0$ requires that
the positive and negative eigenvalues must be paired so as to give a vanishing trace.  Therefore the dimension
$N$ of the underlying Hilbert space must be even, say $N=2K$, and $i_{\rm eff}$ diagonalizes to the form
\begin{equation}\label{eq:diag}
i_{\rm eff}=i {\rm diag}(1,-1,1,-1,...,1,-1)~~~,
\end{equation}
with equal numbers  of eigenvalues $1,\,-1$ along the principal diagonal.

We remark now that the connection between trace dynamics and an emergent quantum theory leads to {\it two
copies} of the quantum theory,  one with a $K$ dimensional Hilbert space on which the effective imaginary unit is $i$, and
the other with a  $K$ dimensional Hilbert space on which the effective imaginary unit is $-i$, corresponding to the two ways
in which $i_{\rm eff}$ can act.  This dichotomy is borne out by calculations \cite{adlerbook} showing that, under suitable approximations,
canonical ensemble averages of products of dynamical variables in trace dynamics can be put into correspondence with
Wightman functions of an  emergent quantum theory.  For a general $N \times N$ matrix $M$, let us denote by $M_{\rm eff}$
the part that commutes with $i_{\rm eff}$, that is $M_{\rm eff}=\frac{1}{2} (M-i_{\rm eff} M i_{\rm eff})$.  Then the
emergent quantum equations take the following form:  For time evolution of effective quantum operators $x_{r\, {\rm eff}}$
 with $x_r$ a $q_r$ or a $p_r$,  we
find the effective Heisenberg equation of motion
\begin{equation}\label{eq:Heis}
\dot x_{r\, {\rm eff}}= \frac{i_{\rm eff}}{\hbar}[H_{\rm eff},x_{r\, {\rm eff}}]~~~.
\end{equation}
For the non-vanishing effective canonical commutators of bosonic degrees of freedom, we find
\begin{equation}\label{eq:bos}
[q_{u \,{\rm eff}},p_{v\, {\rm eff}}]=i_{\rm eff} \hbar \delta_{uv}~~~,
\end{equation}
and for the non-vanishing effective canonical anticommutators of fermionic degrees of freedom,
we find
\begin{equation}\label{eq:ferm}
\{q_{u \,{\rm eff}},p_{v\, {\rm eff}}\}=i_{\rm eff} \hbar \delta_{uv}~~~.
\end{equation}
 On the sector where $i_{\rm eff}=i$, we get the usual
quantum mechanical relations, while on the sector where $i_{\rm eff}=-i$, we get
quantum mechanics with $i$ replaced by $-i$ in the Heisenberg equations of motion
and the canonical commutation/anticommutation relations.  In the concluding chapter of \cite{adlerbook}, and as elaborated
in \cite{adleressay2}, we have suggested that the $-i$ sector is a candidate for the dark matter
sector of the universe.

In applications of trace dynamics to field theory, the discrete index $r$ labeling degrees of freedom
becomes a spatial coordinate label ${\vec x}$.  We show in \cite{adlerbook} that for Lorentz invariant Lagrangians, the operator $\tilde C$
is a Lorentz scalar, and that there is a conserved trace stress energy tensor of the usual form.  We also
show \cite{adlersusy} that the  rigid supersymmetry theories of spin 0, 1/2, and 1 fields, as well as the ``matrix model for M-theory'',
have trace dynamics extensions, whereas \cite{adlerbook} supergravity does not have a trace dynamics extension.

\section{Arguments for the metric being c-number valued in trace dynamics}

We show now that to incorporate gravity into trace dynamics, the metric must be introduced as a c-number or classical field, that
is, as a purely diagonal matrix.  There are a number of independent arguments for this.

\begin{description}
\item [invariant volume]

 Rewriting a flat spacetime theory in curved coordinates requires a spacetime volume element
$dV$ that is invariant under general coordinate transformations.  The usual recipe is $dV = d^4x (^{(4)}g)^{1/2}$, where the scalar density $^{(4)}g$ is given by
$^{(4)}g\equiv - {\rm det} g_{\mu \nu}$ in terms of the metric $g_{\mu \nu}$.  Under a change of coordinates $x_{\mu} \to x_{\mu} (x^{\prime})$,
the scalar density transforms as $(^{(4)}g)^{1/2}\to (^{(4)}g^{\prime})^{1/2} |J|$, with J the Jacobian of the transformation obeying
$|J| d^4x = d^4 x^{\prime}$.   However,
for general operator-valued metric components $g_{\mu \nu}$, the product property of the determinant is lost.  That is,  the determinant
of a matrix, whose elements are the matrix product of an operator valued $g_{\mu \sigma}$ with a c-number
matrix $\partial x^{\sigma}/\partial x^{\prime~\nu}$ is {\it not} the product of the respective determinants of the matrices.  Thus, one
cannot construct the appropriate invariant volume element if $g_{\mu \nu}$ is operator-valued.

\item [obstacle to trace dynamics extension of rigid supersymmetry theories]

If $g_{\mu \nu}$ and $(^{(4)}g)^{1/2}$ are operator valued, the constructions of \cite{adlersusy} for trace dynamics
extensions of rigid supersymmetry theories fail in curved spacetime, because the metric factors do not commute with the matter fields,
and prevent the cyclic permutation of matter fields inside the trace needed to verify supersymmetry.

\item [supergravity]

As already mentioned, for reasons explained in Sec. 3.4 of \cite{adlerbook}, supergravity does not admit a trace dynamics
extension in which the metric and Rarita-Schwinger spinor are operator-valued quantities.

\end{description}

For these reasons, we are led to  introduce the metric into trace dynamics as a c-number field.  There has been considerable
discussion in the literature of whether gravity has to be quantized.  Dyson \cite{dyson} argues that the Bohr-Rosenfeld argument
for quantization of the electromagnetic field does not apply to gravity, and moreover, by a number of examples, shows that  it is hard
(perhaps not possible)  to formulate an experiment
that can detect a graviton.  Dyson also notes that the papers of Page and Geilker \cite {page} and Eppley and Hannah \cite{eppley} , which have been cited to argue that gravity must
be quantized, really only show  that a particular model for classical gravity coupled to quantized matter is inconsistent.  Specifically,
these papers consider the M{\o}ller and Rosenfeld proposal for constructing a semi-classical Einstein equation by writing
$G_{\mu \nu} = -8\pi G\langle \psi|T_{\mu \nu}|\psi \rangle$, and argue that this construction has insurmountable problems when confronted with measurements giving rise to  state
vector reduction.  But,  as Page and Geilker note in their conclusion, while this rules out the semi-classical source postulate, it does not
rule out more complicated forms of classical gravity theories.

To incorporate classical gravity into trace dynamics we proceed as follows.  We start from a flat spacetime trace matter action
\begin{equation}
{\bf S}_{\rm m}= \int dt {\bf L} = \int d^4 x {\rm Tr} {\cal L}(x)~~~,
\end{equation}
with ${\cal L}$ an operator Lagrangian density.  We then generalize this to curved spacetime in the
usual fashion, by introducing a classical metric $g_{\mu \nu}$ and writing
\begin{equation}
{\bf S}_{\rm m}[g]= \int dt {\bf L} = \int d^4 x (^{(4)}g)^{1/2} {\rm Tr} {\cal L}(x;g)~~~,
\end{equation}
with ${\cal L}(x;g)$ the operator Lagrangian density with the classical metric $g_{\mu \nu}$ used
to form covariant derivatives and to contract indices to form  scalars.
The total action will now be
\begin{equation}
{\bf S}_{\rm tot}={\bf S}_{\rm m}[g]+{\bf S}_{\rm g}~~~,
\end{equation}
with the gravitational action ${\bf S}_{\rm g}$ given by
\begin{align}
{\bf S}_{\rm g} = & \frac{1}{16\pi G}{\rm Tr} \int d^4x (^{(4)}g)^{1/2} R\cr
= & \frac{{\rm Tr}(1) }{16 \pi G} \int d^4x (^{(4)}g)^{1/2} R   ~~~,\cr
\end{align}
where $G$ is the gravitational constant and $R$ is the curvature scalar. In the second
line we have used the fact that since the metric is a c-number, $R$ is also a c-number,
so the trace just gives a numerical factor ${\rm Tr}(1)$, which is the dimension of the
underlying Hilbert space. It is convenient now to divide out this factor, by writing
$S_g={\bf S}_g/{\rm Tr}(1)$ and $S_m= {\bf S_m}/{\rm Tr}(1)$.
Varying the metric, we get
\begin{equation}
\delta S_{\rm g}= -\frac{1 }{16 \pi G} \int d^4x (^{(4)}g)^{1/2} G^{\mu \nu}\delta g_{\mu \nu}   ~~~,
\end{equation}
and
\begin{equation}\label{eq:tdef}
\delta S_{\rm m}=-\frac{1}{2} \int d^4 x (^{(4)}g)^{1/2} [{\bf T}^{\mu \nu}/{\rm Tr(1)}] \delta g_{\mu \nu}~~~,
\end{equation}
with ${\bf T}^{\mu \nu}$ the trace stress-energy tensor.
Equating the metric variation of the total action to zero, we get as the trace dynamics gravitational field  equations
\begin{equation}\label{eq:gravfield}
G^{\mu \nu} + \frac {8\pi G}{{\rm Tr}(1)} {\bf T}^{\mu \nu}=0~~~.
\end{equation}
which defining
\begin{equation}
T^{\mu \nu} = \frac {{\bf T}^{\mu \nu}}{{\rm Tr}(1)}
\end{equation}
takes the usual form
\begin{equation}\label{eq:gravfield1}
G^{\mu \nu}+8\pi G {\ T}^{\mu \nu}=0~~~.
\end{equation}
Since ${\bf T}^{\mu \nu}$ obeys the covariant conservation condition
\begin{equation}
\nabla_{\mu} {\bf T}^{\mu \nu}=0~~~,
\end{equation}
Eqs. \eqref{eq:gravfield}-\eqref{eq:gravfield1}  are fully consistent with the gravitational Bianchi identities
\begin{equation}
\nabla_{\mu} G^{\mu \nu}=0~~~,
\end{equation}
Thus, if our conjecture that the underlying equations of  trace dynamics give rise, at the level of thermodynamics
and statistical mechanics, to both quantum theory and state vector reduction, the consistency problems that afflict the
M{\o}ller-Rosenfeld  semi-classical gravity theory are absent in the trace dynamics  framework.

Additionally, we note that
convergence of the partition function $Z$ for the canonical ensemble requires ${\bf H}\geq 0$ over phase space, and if we
were to consider an ensemble translating with velocity $v_i$ with $v_iv^i/c^2 \leq 1$, convergence of the analogous ensemble would require
positivity of ${\bf H} + v_i{\bf P}^i$, with ${\bf P}^i$ the trace momentum.  This positivity requirement is guaranteed if the trace Hamiltonian
and trace three momentum satisfy the ``dominant energy'' condition, which is also the condition needed to prove the positive
energy theorems in relativity.
In the conventional approach to quantum gravity, with a quantum stress-energy tensor as the source of gravity, it has never been clear
why the dominant energy condition should hold after stress-energy tensor regularization.

\section{The matter-induced effective action for gravity}

Even when no particulate matter sources are present, the averaged pre-quantum matter field motions can influence gravitational dynamics.
This is taken into account by defining an induced gravitational action as the action calculated from the average
of the matter field Lagrangian density over the canonical ensemble,
\begin{equation}\label{eq:ginddef}
S_{\rm g; induced}\equiv \int d^4x (^{(4)}g)^{1/2} [{\rm Tr} \langle{\cal L}(x)\rangle_{\rm AV}/{\rm Tr}(1)]~~~,
\end{equation}
where $\langle {\cal L}(x)\rangle_{\rm AV}$ denotes an average over the trace dynamics canonical ensemble $\rho$ of Eq. \eqref{eq:canon},
\begin{equation}\label{eq:avdef}
\langle{\cal L}(x)\rangle_{\rm AV}\equiv  \int d\mu \rho {\cal L}(x) ~~~.
\end{equation}
In more detail, this average is computed as follows.  Writing $x=(x^0,{\vec x})=(t,{\vec x})$, the Lagrangian density  ${\cal L}$ at time $t$
is defined as  a function of the matter fields and their time derivatives. Labeling the matter fields, which can be bosonic or
fermionic,  by an index $a$,   the set of fields are  $q_a(t,{\vec x})$ and their time derivatives are $\dot q_a(t,{\vec x})$.
We  can now rewrite the matter field time derivatives in terms of the corresponding canonical momenta $p_a(t,{\vec x})$  (for
matter gauge fields, this will involve a gauge fixing),  so that the Lagrangian density becomes a function of the
fields and momenta.  Thus at
each fixed time $t$ we can write
\begin{equation}\label{eq:Lqp}
{\cal L}(x)= {\cal L}\big(q_a(t,{\vec x}),p_a(t,{\vec x})\big) ~~~.
\end{equation}
Recall that the canonical ensemble $\rho$ and the phase space measure $d\mu=\prod_a \prod_{\vec x} dq_a(t,{\vec x}) dp_a(t,{\vec x})$  are
time-independent (with $dq$ for a complex matrix $q$ defined in the usual way \cite{adlerbook} as the product of the differentials of the real and imaginary parts of the matrix elements of $q$).
So at fixed time $t$, the average required by Eq. \eqref{eq:avdef} with the Lagrangian density rewritten in the form of Eq. \eqref{eq:Lqp} is now
explicitly defined. Note that the metric $g_{\mu\nu}$ is held fixed in this averaging, which leads to a functional of the metric as stated
in Eq. \eqref{eq:ginddef}.

The assertion that the canonical ensemble $\rho$ is time-independent needs elaboration when in curved spacetime.  It requires that the three quantities
in the exponent of Eq. \eqref{eq:canon}, ${\bf H}$, ${\bf N}$, and $\tilde C$, which are constants of the motion \cite{adlerbook} in flat spacetime, remain constants of the motion in curved spacetime. For ${\bf N}$ and $\tilde C$ this follows from the fact (shown explicitly in Appendix A) that these are charges formed from conserved currents,
which generalize to covariantly conserved currents in curved spacetime. That is,
\begin{align}
\tilde C=&\int d^3x (^{(4)}g)^{1/2} \tilde C^0~~~,\cr
{\bf N} =&\int d^3x (^{(4)}g)^{1/2} {\bf N}^0~~~,\cr
\end{align}
with $\tilde C^{\mu}$ and ${\bf N}^{\mu}$ covariantly conserved four vector currents obeying
$\nabla_{\mu}\tilde C^{\mu}=\nabla_{\mu} {\bf N}^{\mu} =0$.  The usual identity for any
contravariant vector current $V^{\mu}$,
\begin{equation}
\nabla_{\mu} V^{\mu}= (^{(4)}g)^{-1/2}\partial_{\mu}[ (^{(4)}g)^{1/2} V^{\mu}]
\end{equation}
then shows that in curved spacetime, $\tilde C$ and ${\bf N}$ are time-independent.

For the trace Hamiltonian more explanation is needed. In flat spacetime the canonical matter field trace Hamiltonian is defined by
\begin{equation}
 {\bf H}_{\rm m} = \int d^3x{\rm Tr} \sum_{a} p_a(t,{\vec x}) \dot q_a(t,{\vec x}) - {\bf L}~~~.
\end{equation}
Since $1=\delta_0^0$, the natural generalization of this to curved spacetime is
\begin{equation}\label{eq:traceham}
{\bf H}_{\rm m}=\int d^3x (^{(4)}g)^{1/2} {\bf T}_0^0(t,\vec x)~~~,
\end{equation}
with ${\bf T}_{\mu}^{\nu}$ the mixed index trace stress energy tensor.  The fact that we use the mixed
tensor, and not the more customary ${\bf T}^{00}$, will be crucial to the global  Weyl scaling argument that follows.
However, it is well-known that neither of these tensors defines a conserved matter Hamiltonian, because the
energy of the gravitational field must be taken into account. In both cases, it is also known that one can
construct a gravitational stress energy pseudotensor, the Einstein-Dirac \cite{dirac} pseudotensor $t_{\mu}^{\nu}$
in the mixed index case, and the Landau-Lifshitz \cite{landau} pseudotensor $t^{\mu \nu}$ in the upper index case,
that yield conserved quantities.  Specifically, in the mixed index case needed here, $t_{\mu}^{\nu}$ is a function
solely of the metric, constructed so that
\begin{equation}
\partial_{\nu} [(^{(4)}g)^{1/2}( {\bf T}_{\mu}^{\nu}+({\rm Tr}(1)) {t}_{\mu}^{\nu})]=0~~~.
\end{equation}
Thus, when we define
\begin{equation}\label{eq:toten}
{\bf H}= \int d^3x (^{(4)}g)^{1/2} [{\bf T}_0^0(t,\vec x)+({\rm Tr}(1))t_0^0(t,\vec x)]~~~,
\end{equation}
we obtain a Hamiltonian function that is conserved in a general curved spacetime, which can then be used
to construct the canonical ensemble.

We shall not actually
need the detailed form of $t_0^0$, because since it does not depend on the matter fields, it cancels
out of the definition of the canonical ensemble between the numerator in Eq. \eqref{eq:canon} and the
normalizing denominator.  So we can then simply use ${\bf H}_m$ for the Hamiltonian in the canonical
ensemble, and henceforth will drop the subscript $m$. A discussion of the construction of $t_{\mu}^{\nu}$
and its useful properties is given in Appendix B.

\section{Constraints on the form of the induced effective action}

We next address constraints on the structural form of the induced effective action implied by the structure of the
canonical ensemble.  We begin by noting that although $\tilde C$ and ${\bf N}$ are
Lorentz scalars, the trace Hamiltonian ${\bf H}$ is the time component of a four-vector, and so the canonical ensemble
picks out a preferred frame.  We shall make the natural assumption that this preferred frame is the rest frame
of the cosmological background radiation.  However, we shall also assume that there is no other Lorentz violation present,
in particular, we assume that the matter field action in curved spacetime is the usual minimal transcription a
Lorentz invariant flat spacetime action.

Let us consider now a purely spatial general coordinate transformation, which leaves $x^0$ invariant.  Under this transformation, $g_{00}$
transforms as a 3-space scalar, $g_{0i}$ as a three-space covariant vector, and $^{(3)}g\equiv -{\rm det} g_{ij}$ as a three space scalar density.  Expanding
$^{(4)}g$ in a cofactor expansion written in the form
\begin{equation}
^{(4)}g /{^{(3)}g}= g_{00}+g_{0i}D^i~~~,
\end{equation}
we see that $D^i$ transforms as a three-space contravariant vector.  Since the canonical ensemble is a three-space scalar under
purely spatial general coordinate transformations, the induced effective action must share this property, and so must be a
function of the three-space scalars that we can construct from the above quantities and their derivatives, times the invariant volume
element $dV$.  For example, the leading order effective action in an expansion in powers of derivatives of the metric must have the form
\begin{equation}\label{eq:restrict1}
 \Delta S_{\rm g} \equiv S_{\rm g; induced}=  \int d^4x (^{(4)}g)^{1/2} A(g_{00},g_{0i}g_{0j}g^{ij},D^iD^jg_{ij},g_{0i}D^i)~~~.
\end{equation}
with $A(a,b,c,d)$ a general function of its four arguments.

Further restrictions on the form of the induced action come from considering global Weyl scaling invariance.  A detailed study of the Weyl scaling invariance
properties of classical fields in curved spacetime has been given in an important paper by Forger and R\"omer \cite{forger}.
In $n$-dimensional spacetime, they define global Weyl scale transformations of the metric $g_{\mu\nu}$ and the n-bein $e^a_{\mu}$ by the substitutions
\begin{align}
g_{\mu\nu}(x)\to &\lambda^2 g_{\mu\nu}(x) ~~~,\cr
g^{\mu \nu}(x) \to & \lambda^{-2} g^{\mu \nu}(x)~~~,\cr
e^a_{\mu}(x) \to & \lambda e^a_{\mu}(x)~~~,\cr
e_a^{\mu}(x) \to & \lambda^{-1} e_a^{\mu}(x)~~~.\cr
\end{align}
For a  generic matter field $q(x)$ with canonical momentum $p(x)$, the corresponding transformation is
\begin{equation}
q(x) \to \lambda^{-w_q} q(x)~,~~~p(x) \to \lambda^{-w_p} p(x)~~~,
\end{equation}
with
\begin{align}
w_q =& \frac{1}{2}(n-2)~,~~w_p=w_q+2~,~~~q{\rm ~a~scalar~field}~~~(w_q=1~,w_p=3 ~{\rm for~four~dimensions})~~~,\cr
w_q =& \frac{1}{2}(n-1)~,~~w_p=w_q+1~,~~~ q{\rm ~a~Dirac~spinor~field}~~~(w_q=3/2~,w_p=5/2 ~{\rm for~four~dimensions})~~~,\cr
w_q=&w_p = 0~, ~~~q{\rm ~a~Yang-Mills~gauge~field}~~~(w_q=w_p=0 {\rm ~for~four~dimensions})~~~.\cr
\end{align}
Thus, in this scheme, the metric $g_{\mu \nu}$ has Weyl dimension $-2$, and the n-bein $e^a_{\mu}$ has
Weyl dimension $-1$.
The necessity for giving Weyl dimension zero to Yang-Mills fields arises from the fact that the
Yang-Mills field strength $F_{\mu \nu}=\nabla_{\mu} A_{\nu}-\nabla_{\nu} A_{\mu} + A_{\mu} \times A_{\nu}$
involves both linear and quadratic terms in the gauge potential $A_{\mu}$.
 (Forger and R\"omer do not consider  $U(1)$ gauge
fields, for which one could consistently assign a scale dimension of $w_q=\frac{1}{2}(n-4)$ in $n$ dimensions; however,
from a grand unification point of view, $U(1)$ fields arise from Yang-Mills fields, and are not present in the
original matter action.)

Forger and R\"omer study the global Weyl transformation properties of standard actions for massless spin 0, 1/2, and 1 matter fields in
curved spacetime and find the following results (which will be derived in Appendix A):

\begin{enumerate}

\item  The massless spin 0 actions, both ``improved'' with an additional term  $b(n) R \phi^2$, and ``minimal'' without this term, are
 globally Weyl invariant {\it off-shell} (without use of the equations of motion)  in $n$ dimensions.

\item The massless Dirac spinor action is globally  Weyl invariant off-shell in
$n$ dimensions.

\item The Yang-Mills gauge field action is globally Weyl invariant  off-shell only in $n=4$ dimensions.

\end{enumerate}

In order to study the Weyl scaling  properties of the canonical ensemble, we must additionally know the Weyl properties
of the trace stress-energy tensor ${\bf T}_0^0$, as well as of $\tilde C$ and ${\bf N}$.  In Appendix A we also
give the formulas for $T_{\mu \nu}$ given by Forger and R\"omer for classical fields. (These are converted to
formulas for ${\bf T}_{\mu \nu}$ by reinterpreting the fields as matrix valued and symmetrizing coupling terms
where there are factor-ordering ambiguities. However, since the Weyl scaling factor $\lambda$ for a classical metric
is necessarily classical, the Weyl invariance calculations for both the action and the stress-energy tensor
are the same for both  the classical models in Appendix A and their trace dynamics transcriptions.)
In $n=4$ dimensions, we find the following  Weyl scaling properties:

\begin{enumerate}

\item From the global Weyl invariance of the matter field actions, and the definition  of Eq. \eqref{eq:matteraction}, we
deduce that $(^{(4)}g)^{1/2} T_{\mu}^{\nu}$ is off-shell globally Weyl invariant for the massless scalar, massless
Dirac, and Yang-Mills gauge fields.

\item  For matrix-valued scalar, Yang-Mills, and Dirac spinor fields, as needed to construct the current
${\tilde C}^{\mu}$, we find off-shell that $(^{(4)}g)^{1/2}{\tilde C}^{\mu}$ is globally Weyl invariant.

\item For matrix-valued Dirac spinor fields, as needed to construct the trace fermion number current ${\bf N}^{\mu}$, we find off-shell that
$(^{(4)}g)^{1/2}{\bf N}^{\mu}$ is globally Weyl invariant.
\end{enumerate}

To summarize these results, {\it all} of the on-shell conserved quantities used to form the canonical ensemble are {\it off-shell} invariant
under global  Weyl scalings. This means that they are globally Weyl invariant over the entire phase space that is integrated over in the
canonical ensemble.  Since the Weyl scaling factors cancel between the phase space measure factors $d\mu$ in Eq. \eqref{eq:canon},
and since the matter action is globally Weyl scaling invariant, we learn that the matter induced gravitational effective action defined in
Eq. \eqref{eq:ginddef} must be globally Weyl invariant.  Since $D^i$ has Weyl scaling weight 0, this allows us to further
restrict the functional form given in Eq. \eqref{eq:restrict1} to read
\begin{equation}\label{eq:restrict2}
\Delta  S_{\rm g}= \int d^4x (^{(4)}g)^{1/2}(g_{00})^{-2} A \big(g_{0i}g_{0j}g^{ij}/g_{00},D^iD^jg_{ij}/g_{00},g_{0i}D^i/g_{00}\big)~~~.
\end{equation}
with $A(x,y,z)$ a general function of its three arguments. We remark that this result excludes a cosmological constant term in the
induced gravitational action, which would correspond to an action
\begin{equation}
S_{\rm cosmological~constant}\propto\int d^4x (^{(4)}g)^{1/2}~~~
\end{equation}
that is not globally Weyl scaling invariant. (Thus, we have given here a corrected version of the argument which we initially
attempted in \cite{adleressay1}).  We also remark that in the important case of metrics for which $g_{0i}=g^{0i}=D^i=0$, Eq. \eqref{eq:restrict2}
greatly simplifies to read
\begin{equation}\label{eq:restrict3}
\Delta S_{\rm g}=A_0 \int d^4x (^{(4)}g)^{1/2}(g_{00})^{-2}~~~,
\end{equation}
where $A_0=A(0,0,0)$ is a constant factor. Similarly, when $g_{0i}$ and $D^i$ are effectively small, as for the metrics for slowly
rotating bodies, we can expand $A(x,y,z)$ to first order in its arguments, giving the effective action
\begin{equation}\label{eq:restrict4}
\Delta S_{\rm g}= \int d^4x (^{(4)}g)^{1/2}(g_{00})^{-2} [A_0 + (g_{00})^{-1}
 (B_1 g_{0i}g_{0j}g^{ij}+B_2 D^iD^jg_{ij}+B_3g_{0i}D^i)]~~~.
\end{equation}

\section{Rules for use of the frame-dependent effective action}

 When particulate matter (baryonic matter, dark matter, and radiation)  is present, with action $S_{\rm pm}$,
the total action that we have obtained is
\begin{equation}\label{eq:stotal}
S_{\rm total} = S_g + \Delta S_{\rm g} + S_{\rm pm}~~~.
\end{equation}
The familiar actions $S_g$ and $S_{\rm pm}$ are general coordinate transformation scalars, but the induced action
$\Delta S_{\rm g}$ is frame dependent, and as we have seen is only invariant under the subset of general coordinate
transformations that act on the spatial coordinates $\vec x$, but leave the time coordinate $t$ invariant.  As a result,
the spacetime stress energy tensor obtained by varying $\Delta S_{\rm g}$ with respect to the full metric $g_{\mu \nu}$
will not satisfy the covariant conservation condition, and thus cannot be used as a source for the full spacetime
Einstein equations.  However, it is perfectly consistent to use $\Delta S_{\rm g}$ as the source for the spatial
components of the Einstein tensor $G^{ij}$ in the preferred rest frame of the canonical ensemble, which we have
assumed to be the rest frame of the cosmological background radiation.  Thus we get the following rules:

\begin{enumerate}

\item  The spatial components $G^{ij}$ of the Einstein equations are obtained by varying $S_{\rm total}$ of Eq. \eqref{eq:stotal} with
respect to the spatial components $g_{\ij}$ of the metric tensor, giving the gravitational field equations
\begin{equation}\label{eq:split}
G^{ij}+ 8\pi G(\Delta T^{ij}+ T^{ij}_{\rm pm})=0~~~,
\end{equation}
with $T^{ij}_{\rm pm}$ the spatial components of the usual particulate matter stress-energy tensor $T^{\mu \nu}_{\rm pm}$,
which is covariantly conserved, and with $\Delta T^{ij}$ given by
\begin{equation}\label{eq:tdef}
\delta \Delta S_{\rm g}=-\frac{1}{2} \int d^4 x (^{(4)}g)^{1/2} \Delta T^{ij } \delta g_{ij}~~~.
\end{equation}

\item The components of the Einstein tensor $G^{0i}=G^{i0}$ and $G^{00}$ are obtained from the Bianchi identities, with
$G^{ij}$ as input, and from them we can infer the conserving extensions $\Delta T^{i0}$ and $\Delta T^{00}$ of the induced
gravitational stress energy tensor.  Equivalently, we can infer these by imposing the covariant conservation condition
on the full induced tensor $\Delta T^{\mu \nu}$, with $\Delta T^{ij}$ as input.

\item  With this interpretation, comparing Eq. \eqref{eq:split} with Eq. \eqref{eq:gravfield1}, we see that we have defined a splitting
of the trace matter stress energy tensor into a part  $\Delta T^{\mu \nu}$ that arises from the hidden averaged motions of the pre-quantum matter
fields, and a part $T^{\mu \nu}_{\rm pm}$ that arises from the observable particulate matter,
\begin{equation}
T^{\mu \nu} = \frac {{\bf T}^{\mu \nu}}{{\rm Tr}(1)}= \Delta T^{\mu \nu} + T^{\mu \nu}_{\rm pm}~~~.
\end{equation}
\end{enumerate}

These rules have an analog in statistical mechanics and condensed matter theory, where there is a large literature showing
how to obtain ``conserving approximations'' when the full equations of motion are truncated or are averaged over certain
dynamical variables.  Fortunately, the general relativity case just described is simpler.  We shall see that for certain
metrics of particular interest, such as the Robertson-Walker cosmological metric, and the static spherically
symmetric metric, it is easy to write down the conserving extension of $\Delta T^{ij}$.

\section{Application to Robertson-Walker cosmology}

The standard $\Lambda CDM$ model of cosmology, which is in excellent agreement with observational data from the
WMAP and Planck satellites, is based on the Robertson-Walker line element
\begin{equation}
ds^2=dt^2 - a(t)^2\left[\frac{dr^2}{1-kr^2} + r^2 (d\theta^2 + \sin^2\theta d\phi^2)\right]~~~,
\end{equation}
corresponding to the metric components
\begin{equation}
g_{00}=1~,~~g_{rr}=-a(t)^2/(1-k r^2)~,~~g_{\theta \theta}=-a(t)^2 r^2~,~~g_{\phi \phi}=-a(t)^2 r^2 \sin^2\theta~~~.
\end{equation}
Since $g_{0i}=g_{i0}=0~,~~D^i=0$, we can use the simplified form of the induced action given in Eq. \eqref{eq:restrict3}.
Substituting $g_{00}=1$, we get
\begin{equation}\label{eq:cosmo}
\Delta S_{\rm g}=A_0 \int d^4x (^{(4)}g)^{1/2}~~~.
\end{equation}
Varying the spatial components $g_{ij}$ of the metric, while taking $\delta g_{00}=\delta g_{0i}=0$,
and using $\delta (^{(4)}g)^{1/2}=\frac{1}{2} (^{(4)}g)^{1/2} g^{\mu \nu}\delta g_{\mu \nu}$,
we find from Eq. \eqref{eq:tdef} that the spatial components of $\Delta T^{ij}$ are given by
\begin{equation}
\Delta T^{ij}=-A_0 g^{ij}~~~.
\end{equation}
The conserving extension of the induced gravitational stress-energy tensor for this case is obviously given by
\begin{equation}
\Delta T^{\mu \nu}=-A_0 g^{\mu \nu}~~~,
\end{equation}
and we see that for a homogeneous, isotropic cosmological metric, the induced term has {\it exactly}
the structure of a cosmological constant!  Assuming that there is no ``bare'' cosmological constant, the
induced term is to be identified with the observed cosmological constant. In this interpretation
{\it the so-called ``dark energy'' is the
energy associated with the hidden motions of the pre-quantum matter fields}, and is strictly constant over
the course of cosmic evolution, even as the matter sector undergoes  phase transitions associated
with successive stages of spontaneous symmetry breaking.  Comparing with the standard form of the Einstein
equations in the presence of a cosmological constant $\Lambda$,
\begin{equation}
G^{\mu \nu}+\Lambda g^{\mu \nu} +8\pi G T_{\rm pm}^{\mu \nu}=0~~~,
\end{equation}
we identify the constant $A_0$ in Eq. \eqref{eq:restrict3} as
\begin{equation}\label{eq:avalue}
A_0= -\frac{\Lambda}{8 \pi G}~~~.
\end{equation}
Using the relation $\Lambda = 3 H_0^2 \Omega_{\Lambda}$ between
$\Lambda$, the Hubble constant $H_0$ and the cosmological fraction $\Omega_{\Lambda}$,
we get the alternative expression
\begin{equation}
A_0=-\frac {3 H_0^2 \Omega_{\Lambda} } { 8 \pi G}~~~.
\end{equation}
We emphasize that we are inferring the value of $A_0$ from the experimentally observed cosmological constant, and so have
{\it not} given an explanation of why $A_0$ is so small compared to the scale set by the Planck mass.
\section{A first look at the static, spherically symmetric metric}

The standard form for the static, spherically symmetric line element is
\begin{equation}
ds^2=B(r) dt^2 - A(r) dr^2 - r^2( d\theta^2 + \sin^2 \theta d\phi^2)~~~,
\end{equation}
corresponding to the metric components
\begin{equation}
g_{00}=B(r)~,~~g_{rr}=-A(r)~,~~g_{\theta \theta}= -r^2~,~~g_{\phi \phi}= -r^2 \sin^2\theta~~~.
\end{equation}
Again we have $g_{0i}=g_{i0}=0~,D^i=0$, so we can again use the simplified form of the
induced action given in Eqs. \eqref{eq:restrict3} and \eqref{eq:avalue}.
Substituting $g_{00}=B(r)$, we get
\begin{equation}\label{eq:spher}
\Delta S_{\rm g}=-\frac{\Lambda}{8 \pi G} \int d^4x (^{(4)}g)^{1/2} B(r)^{-2}~~~.
\end{equation}
Again by varying the spatial components $g_{ij}$ of the metric, while taking $\delta g_{00}=\delta g_{0i}=0$,
we find from Eqs. \eqref{eq:tdef} and \eqref{eq:spher}  that the spatial components $\Delta T^{ij}$
are given by
\begin{equation}
\Delta T^{ij}=\frac{\Lambda}{8 \pi G} g^{ij}/B(r)^2~,~~~ \Delta T_{ij} = \frac{\Lambda}{8 \pi G}g_{ij}/B(r)^2~~~.
\end{equation}
and the Einstein equations for $G_{rr}$ and $G_{\theta \theta}$ are modified to read
\begin{align}
&G_{rr}- \frac {\Lambda A(r)}{ B(r)^2}=0~~~,\cr
&G_{\theta \theta} -\frac {\Lambda r^2} { B(r)^2}=0~~~,\cr
\end{align}
with the equation for $G_{\phi \phi}$ proportional to
that for $G_{\theta \theta}$.  Although $\Lambda$ is very small, we see that near
the horizon of a Schwarzschild black hole, where the unperturbed solution is $A(r)^{-1}=B(r)=1-r_S/r$ (with
$r_S$ the Schwarzschild radius), the induced term becomes infinite and so may have
a significant effect on the horizon structure, which we plan to study.

From the expressions for $G_{tt}$, $G_{rr}$, and $G_{\theta \theta}$, with $^{\prime}$ denoting $d/dr$,
and with $A\equiv A(r)$ and $B \equiv B(r)$ in Eqs. \eqref{eq:spher1} -- \eqref{eq:spher2},
\begin{align}\label{eq:spher1}
G_{tt}=&\frac{B}{rA} \left[ -\frac{A^{\prime}}{A}+ \frac{1}{r} (1-A)\right]~~~,\cr
G_{rr}=&-\frac{B^{\prime}}{rB} + \frac{1}{r^2} (A-1)~~~,\cr
G_{\theta \theta}= -&\frac{r^2}{2A}\left[\frac{B^{\prime \prime}}{B} - \frac{B^{\prime}}{2B}
\left(\frac{A^{\prime}}{A}+\frac{B^{\prime}}{B}\right)+\frac{1}{r}\left(-\frac{A^{\prime}}{A}+\frac{B^{\prime}}{B} \right) \right]~~~,\cr
\end{align}
we find the linear relation (the Bianchi identity)
\begin{equation}\label{bianchi}
G_{rr}^{\prime} -\frac{2 A}{r^3} G_{\theta \theta} +\left( \frac{B^{\prime}}{2B}+\frac{2}{r}-\frac{A^{\prime}}{A} \right) G_{rr}
+\frac{A B^{\prime}}{2B^2}G_{tt}=0~~~.
\end{equation}
When $G_{\mu \nu}$ is replaced in this equation by the covariantly conserved $\Delta T_{\mu \nu}$ it must also be satisfied,  so for the
conserving extension $\Delta T_{tt}$ of $\Delta T_{rr}$ and $\Delta T_{\theta \theta}$ we find
\begin{align}\label{eq:spher2}
\Delta T_{tt}=&N/D~~~,\cr 
N=&\Delta T_{rr}^{\prime} -\frac{2 A}{r^3} \Delta T_{\theta \theta} +\left( \frac{B^{\prime}}{2B}+\frac{2}{r}-\frac{A^{\prime}}{A} \right) \Delta T_{rr}~~~,\cr
D=&-\frac{A B^{\prime}}{2B^2}~~~.\cr
\end{align}
Using $\Delta T_{rr}$ and $\Delta T_{\theta \theta}$  from Eq. (58), this gives as the modified equation for $G_{tt}$
\begin{equation}
G_{tt}-\frac {3 \Lambda}{B}=0~~~.
\end{equation} 

\section{Some remarks}
In conclusion we make some remarks, first on subtleties of our derivations, and then on speculations and possible future directions.

\subsection{Remarks on the derivations}

 (1)~~Our calculation is a form of a ``fast-slow'' calculation, in which ``fast'' degrees of freedom (in our case, the pre-quantum matter) are averaged
 to get effective equations for ``slow'' ones, in our case, the metric.  In usual applications of this method, one averages the Hamiltonian instead
 of the action.  When the action and Hamiltonian have the form (using subscripts $f$ and $s$ to label the fast and slow degrees of freedom)
 \begin{align}
 S=&T-V=T(\dot q_f)+T(\dot q_s) -V(q_f,q_s)~~~,\cr
 H=&T+V=T(p_f)+ T(p_s) +V(q_f,q_s)  ~~~,\cr
 \end{align}
 and one averages over a normalized weighting of either the form $\rho(q_f,p_f)$, or the factorized form $\rho(p_f) \eta(q_f,q_s) $ (which includes
 the thermal ensemble for this system), the
 averages of $T(p_f)$ and $T(\dot q_f)$ are constants which do not contribute to the averaged Euler-Lagrange and  Hamilton equations
 for the slow variables. The averaged action and the averaged Hamiltonian then give the same equations of motion.  However, when the fast kinetic terms depend on slow variables, and so have the form $T(\dot q_f, q_s)$ or $T(p_f,q_s)$, when averaged these give  extra potential-like
 terms for the slow variables, and the two averaging procedures are not manifestly equivalent.  This corresponds to the gravitational case that we have studied, where the matter kinetic terms depend on the metric. Since the action form of gravitation is much simpler than the Hamiltonian form, we have chosen
 to average the action.

 (2)~~We have focussed on global Weyl invariance properties, which are enough to restrict the leading terms in the induced gravitational effective
 action in an expansion in powers of derivatives of the metric.  As discussed in Appendix A, the ``improved'' scalar,  Dirac, and Yang-Mills actions and Hamiltonians
in $n=4$ dimensions are also invariant under local Weyl transformations with $\lambda = \lambda(x)$, and therefore under time independent
transformations with $\lambda=\lambda(\vec x)$, which can still be scaled out of the canonical ensemble integration measure $d\mu$. For Dirac and
Yang-Mills fields, $(^{(4)}g)^{1/2} {\cal L}$ is also locally Weyl invariant, while in the scalar case, $(^{(4)}g)^{1/2} {\cal L}$ is locally
invariant up to a total derivative, which does not contribute to the action.
However,  in the scalar case, since $p_{\phi}=g^{0\mu}\partial_{\mu} \phi= g^{00} \partial_0 \phi
+ g^{0i} \partial_i \phi$, the local scaling properties of $p_{\phi}$ and $\phi$ are consistent only when $\lambda=\lambda(\vec x)$
 and when the metric is specialized to $g^{0i}=0$.  These results place no additional restrictions on the leading non-derivative effective action terms, but can be used to place scaling restrictions on the terms in the effective action that depend on derivatives of the metric.

\subsection{Speculations and possible future directions}
 (1)~~In general there will be induced corrections to the $R$ term in the gravitational action, with a  coefficient $C$  of dimension
 $[{\rm mass}]^2$, whereas the coefficient $A$ of the leading term in the derivative expansion has dimension $[{\rm mass}]^4$.  If we make the
 naive estimate $C \sim A^{1/2}$, then the size of the correction to the $R$ action relative to the Einstein-Hilbert action will be of order $G A^{1/2} \sim H_0 G^{1/2} \sim 10^{-60}$,
 that is the ``induced gravitational'' action is much too small to serve as the gravitational action by itself.  So a fundamental $R$ action is needed.
 This suggests studying trace dynamics generalizations of various extended supergravity theories as a possible way of unifying the matter and gravity sectors.

 (2)~~  Because the parameter $\tau$ with dimension of $[{\rm mass}]^{-1}$ appears in the canonical ensemble, there are two constants with dimension
 of  $[{\rm mass}]^{-1}$  present, $G^{-1/2}$ and $\tau$.  We suggest that these can be related by imposing an initial condition of zero total energy $\langle {\bf H}\rangle_{\rm AV}=0$, with ${\bf H}$ including the gravitational energy as in Eq. \eqref{eq:toten}. It well known \cite{muk} that in a Newtonian universe,
 a spatially flat universe with $\Omega=1$ corresponds to zero total energy, with the kinetic energy of matter balanced by the negative potential energy of
 its gravitational attraction.  We show in Appendix B that this statement has a general relativistic analog:  In a spatially flat universe (Robertson-Walker with $k=0$), using Cartesian coordinates, the Einstein-Dirac pseudotensor takes the locally uniform value
 \begin{align}\label{eq:spatflat}
  t_{0~{\rm ED}}^0=& -\frac{3}{8\pi G} \left(\frac{\dot a} {a}\right)^2=-\rho_{\rm tot}~~~,\cr
 t_{0~{\rm ED}}^j=&0~~~,\cr
 \end{align}
 with $\rho_{\rm tot}$ the total matter contribution to the energy density coming from the sum of the induced gravitational term $\Delta T_0^0$ and
  and the particulate matter term $T_{0~{\rm pm}}^0$  in the Einstein equations. Thus in a general relativistic sense as well, the observation of $\Omega=1$  corresponds to a zero energy condition.

\section{Acknowledgements}

I wish to thank Gerhard Gr\"ossing for inviting me to be a Keynote Speaker at the second international
conference on ``Emergent Quantum Mechanics'', EmQM13, to be held in Vienna October 4th-6th,  2013.  This invitation
prompted me to return to the subject addressed in my book \cite{adlerbook}, on which I had talked at the first
conference in Vienna two years ago, to try to address some of the problems that I had left unresolved.
I am grateful to Freeman Dyson for sending me a pre-publication copy of his Poincar\'e Prize Lecture \cite{dyson},
and for email correspondence about the issues he discusses in it.  I also wish to thank Angelo Bassi for reading
the manuscript. Final revisions were supported in part by the National Science Foundation under
Grant No. PHYS-1066293 and the hospitality of the Aspen Center for Physics.

\appendix

\section{Notational conventions and formulas for matter field actions and conserved quantities derived from them}
\subsection{Notational conventions}

Since the books on gravitation and cosmology that we have consulted use many different conventions, we summarize
our notational conventions here. They follow the conventions of the book of Parker and Toms \cite{parker} and the paper of Forger
and R\"omer \cite{forger}.

(1)~~ The Lagrangian in flat spacetime is $L=T-V$, with $T$ the kinetic energy and $V$ the potential energy, and
the flat spacetime Hamiltonian is $H=T+V$.

(2)~~ We use a $(1,-1,-1,-1)$ metric convention, so that in flat spacetime, where the metric is denoted by $\eta_{\mu \nu}$, the various 00 components of
the stress energy tensor $T_{\mu \nu}$ are equal, $T_{00}=T_{0}^{0}=T^{00}$.

(3)~~The affine connection, curvature tensor, contracted curvatures, and the Einstein tensor, are
given by
\begin{align}
\Gamma^{\lambda}_{\mu \nu} =&\frac{1}{2}g^{\lambda \sigma}(g_{\sigma \nu,\, \mu}+g_{\sigma \mu,\, \nu}- g_{\mu \nu,\, \sigma})~~~,\cr
R^{\lambda}_{\tau \mu \nu}=& \Gamma^{\lambda}_{~\tau \mu,\, \nu}-\Gamma^{\lambda}_{~\tau \nu,\, \mu} +{\rm quadratic ~terms ~in~} \Gamma  ~~~,\cr
R_{\mu \nu}=&R^{\lambda}_{~\mu \lambda \nu}=-\Gamma^{\lambda}_{\mu \nu, \, \lambda}+{\rm other~terms}~~~,\cr
R=&g^{\mu \nu} R_{\mu \nu}~~~,\cr
G_{\mu \nu}=&R_{\mu \nu}-\frac{1}{2} g_{\mu \nu} R ~~~.\cr
\end{align}

(4)~~ The gravitational action, with cosmological constant $\Lambda$, and its variation with respect to the metric $g_{\mu \nu}$ are
\begin{align}
S_{\rm g} = & \frac{1}{16\pi G} \int d^4x (^{(4)}g)^{1/2} (R-2\Lambda)~~~,\cr
\delta S_{\rm g}=& -\frac{1 }{16 \pi G} \int d^4x (^{(4)}g)^{1/2} (G^{\mu \nu}+\Lambda g^{\mu \nu})\delta g_{\mu \nu}~~~.\cr
\end{align}

(5)~~The matter action and its variation with respect to the metric $g_{\mu \nu}$ are
\begin{align}\label{eq:matteraction}
S_{\rm m}=& \int dt  L = \int d^4 x (^{(4)}g)^{1/2}  {\cal L}(x)~~~,\cr
\delta S_{\rm m}=&-\frac{1}{2} \int d^4 x (^{(4)}g)^{1/2}  T^{\mu \nu}  \delta g_{\mu \nu}~~~.\cr
\end{align}

(6)~~The Einstein equations are
\begin{equation}
G^{\mu \nu}+ \Lambda g^{\mu \nu} +8 \pi G T^{\mu \nu}=0~~~.
\end{equation}

(7)~~The gravitational covariant derivative that leaves the metric invariant is denoted by $\nabla_{\mu}$, an ordinary partial derivative by $\partial_{\mu}$, and a covariant
derivative with respect to both the metric and gauge fields, by $D_{\mu}$

(8)~~ For uniformly distributed matter in the rest frame of the Robertson-Walker metric,
the stress energy tensor is
\begin{equation}
T^{\mu \nu}= (p+\rho) u^{\mu}u^{\nu} -p g^{\mu \nu}~~~,
\end{equation}
with $p$ the pressure, $\rho$ the mass density, and $u^0=1~,~~u^i=0$.

\subsection{Formulas for matter field actions and conserved quantities derived from them}

To calculate local Weyl scaling properties, it is convenient to write $\lambda(x)=\exp(\omega(x))$ and to study
the effect of an infinitesimal $\omega$.  The calculations given in \cite{forger}
show that under Weyl scaling of the metric and the matter fields in $n=4$ dimensions,  the matter Lagrangian densities all obey
$\delta_{\omega} {\cal L}=-4 \omega {\cal L}$ off-shell, which since $\delta_{\omega} (^{(4)}g)^{1/2} =4 \omega (^{(4)}g)^{1/2}$, implies that
\begin{equation}\label{eq:localinv}
\delta_{\omega} [(^{(4)}g)^{1/2} {\cal L}]=0
\end{equation}
off-shell.  This in turn  implies the invariance of the
corresponding action integral
\begin{equation}\label{eq:actioninv}
\delta_{\omega} \int d^4x (^{(4)}g)^{1/2} {\cal L}=0~~~.
\end{equation}

The matter Lagrangian densities used in \cite{forger}, for which these properties hold, are as follows:

(1)~~The ``improved'' or ``modified'' scalar field Lagrangian density (with $\Box=\nabla^{\mu} \nabla_{\mu}$),

\begin{equation}
{\cal L}_{\rm  scalar}= -\frac{1}{2}\phi \Box \phi - K \phi^4 +\frac{1}{12}R \phi^2~~~,
\end{equation}

(2)~~ the Yang-Mills gauge field Lagrangian density [with (~,~) the internal index inner product],
\begin{equation}
{\cal L}_{\rm gauge}=-\frac{1}{4} g^{\mu \kappa} g^{\nu \lambda} (F_{\mu \nu}, F_{\kappa \lambda})~~~,
\end{equation}

(3)~~the Dirac spinor field Lagrangian density (with $\bar \psi = \psi^{\dagger} {\underline \gamma}_0$ and $\gamma_{\mu} =e^a_{\mu} {\underline \gamma}_a$, with
$\underline \gamma_a$ the flat spacetime gamma matrices),
\begin{equation}
{\cal L}_{\rm spinor}=\frac{i}{2} g^{\mu \nu} \bar \psi \gamma_{\mu} \overleftrightarrow{\nabla}_{\nu} \psi~~~,
\end{equation}

(4)~~the usual renormalizable interaction Lagrangian densities  ${\cal L}_{\rm interaction}$ for gauge fields coupling to scalar and spinor fields (obtained by replacing $\nabla_{\mu} \to D_{\mu}$) and for  Dirac spinors  with Yukawa couplings to scalars,

We note that for the alternative form of the modified scalar Lagrangian density, which differs only by a total derivative that does not contribute to
the action,
\begin{equation}
{\cal L}^{\prime}_{\rm  scalar}= \frac{1}{2}g^{\alpha \beta} \partial_{\alpha}\phi \partial_{\beta} \phi - K \phi^4 +\frac{1}{12}R \phi^2~~~,
\end{equation}
we find
\begin{equation}
\delta_{\omega}[(^{(4)}g)^{1/2} {\cal L}^{\prime}_{\rm  scalar}]= -\frac{1}{2}\partial_{\kappa}[(^{(4)}g)^{1/2}\phi^2 \partial^{\kappa} \omega]~~~.
\end{equation}
Verifying these local Weyl scaling properties requires considerable calculation.  Local Weyl scaling implies global Weyl scaling, which
is obvious by inspection of the above formulas.

By varying the above Lagrangians with respect to the metric, which again requires lengthy calculations, Forger and R\"omer \cite{forger} calculate
formulas for the corresponding stress-energy tensors $T_{\mu \nu}$. Raising the index $\nu$, these become

(1)~~The ``modified'' scalar field stress-energy tensor
\begin{equation}
T_{\mu~{\rm scalar}}^{\nu}=\partial_{\mu}\phi \partial^{\nu}\phi - \delta_{\mu}^{\nu} {\cal L}^{\prime}_{\rm scalar}
+  \frac{1}{6}(\delta_{\mu}^{\nu} \Box -g^{\nu \alpha} \nabla_{\mu}\nabla_{\alpha} + R_{\mu}^{\nu}) \phi^2~~~,
\end{equation}

(2)~~ the Yang-Mills gauge field stress-energy tensor
\begin{equation}
T_{\mu~{\rm gauge}}^{\nu}=-g^{\kappa \lambda} (F_{\mu \kappa},F^{\nu}_{~\, \lambda})- \delta_{\mu}^{\nu} {\cal L}_{\rm gauge}~~~,
\end{equation}

(3)~~the Dirac spinor stress energy-tensor
\begin{equation}
T_{\mu~{\rm spinor}}^{\nu}=\frac{i}{4} g^{\nu \alpha}(\bar \psi \gamma_{\mu} \overleftrightarrow{\nabla}_{\alpha} \psi
+\bar \psi \gamma_{\alpha} \overleftrightarrow{\nabla}_{\mu} \psi) - \delta_{\mu}^{\nu} {\cal L}_{\rm spinor}~~~,
\end{equation}

(4)~~the usual contribution to the stress-energy tensor arising from the interaction Lagrangian densities
\begin{equation}
T_{\mu~{\rm interaction}}^{\nu}=-\delta_{\mu}^{\nu} {\cal L}_{\rm interaction}~~~.
\end{equation}

By direct calculation,  we have verified that the mixed component tensors $T_{\mu}^{\nu}$ for the modified scalar, gauge field, and spinor
cases, satisfy the satisfy the following local
Weyl scaling conditions,
\begin{equation}
\delta_{\omega}[ (^{(4)}g)^{1/2} T_{\mu~ {\rm scalar}}^{\nu}]=
\delta_{\omega}[ (^{(4)}g)^{1/2} T_{\mu~ {\rm gauge}}^{\nu}] =\delta_{\omega}[ (^{(4)}g)^{1/2} T_{\mu~ {\rm spinor}}^{\nu}]=0~~~.
\end{equation}
Thus, in all three cases, we learn that the three space integral
\begin{equation}
\int d^3 x (^{(4)}g)^{1/2} T_{\mu}^{\nu}
\end{equation}
is Weyl scale invariant for general time-independent but space dependent $\omega(\vec x)$. The global Weyl invariance specialization of
these results can again be read off from the expressions for the stress-energy tensors without detailed calculation.

These local invariance results for the stress-energy tensor can also be deduced from the local Weyl variation  of the action by the
following argument.   Since the order of variations can be interchanged, we have
\begin{equation}\label{eq:interchange}
\delta_{\omega} \delta_{g^{\mu \nu}}=\delta_{g^{\mu \nu}}\delta_{\omega} ~~~
\end{equation}
Applying the right hand side to the product of $(^{(4)}g)^{1/2} $ with the Lagrangian density,  and using Eq. \eqref{eq:localinv} we get
\begin{equation}\label{eq:inv1}
\delta_{g^{\mu \nu}}\delta_{\omega}  [(^{(4)}g)^{1/2} {\cal L}]=0~~~.
\end{equation}
Now in general, the metric variation of $[(^{(4)}g)^{1/2} {\cal L}]$ has the form
\begin{equation}\label{eq:metricvar}
\delta_{g^{\mu \nu}} [(^{(4)}g)^{1/2} {\cal L}]= \frac{1}{2} (^{(4)}g)^{1/2}  T_{\mu \nu}\delta g^{\mu \nu} + \partial_{\kappa}\Sigma^{\kappa}(\delta g^{\mu \nu})~~~,
\end{equation}
with the second term consisting of total derivatives that are discarded after integrating over $d^4x$.  So from Eqs. \eqref{eq:interchange} -- \eqref{eq:metricvar}
 we find
\begin{equation}
\delta_{\omega}  [(^{(4)}g)^{1/2}  T_{\mu \nu}\delta g^{\mu \nu} + 2 \partial_{\kappa}\Sigma^{\kappa}(\delta g^{\mu \nu})]=0~~~.
\end{equation}
Since $\delta g^{\mu \nu}$ is arbitrary, we can take it as
\begin{equation}\label{eq:deltagsp}
\delta g^{\mu \nu} = \frac{1}{2}(g^{\nu \alpha} \eta^{\mu} + g^{\mu \alpha} \eta^{\nu})\xi_{\alpha} ~~~,
\end{equation}
with $\xi_{\alpha}$ and $\eta^{\alpha}$ constant four vectors. We then learn
\begin{equation}
\delta_{\omega}[ (^{(4)}g)^{1/2} T_{\mu}^{\alpha}] \xi_{\alpha}\eta^{\mu}=-2 \delta_{\omega}\partial_{\kappa} \Sigma^{\kappa}(\xi,\eta)~~~.
\end{equation}
In the gauge field and Dirac spinor cases the total derivative term $\Sigma^{\kappa}$ vanishes (for the spinor, this requires a lengthy calculation
given in \cite{forger}), and so using the fact that $\xi$ and $\eta$ are arbitrary, we learn that
\begin{equation}\label{eq:scalar}
 \delta_{\omega}[(^{(4)}g)^{1/2} T_{\mu}^{\alpha}]=0~~~,
\end{equation}
which is the result obtained by direct calculation. In the scalar case,  several integrations by parts are needed to get from the
variation of the action to the stress-energy tensor, so $\Sigma^{\kappa}$ is nonzero. To get the detailed form of $\delta_{\omega} \Sigma^{\kappa}$, one  must  calculate the surface term $\Sigma^{\kappa}$, which we have done as an independent check on the scalar case results stated above, but which involved
considerable effort.

We stress again that all of the above Weyl scaling results are valid ``off-shell'', that is without use of the equations of motion.  The main
focus of Forger and R\"omer \cite{forger} was not on Weyl scaling of the stress-energy tensor, but rather on the connection between scale invariance and vanishing of the
trace $T_{\mu}^{\mu}$ of the stress-energy tensor.  Here the equations of motion are used in their Theorem 5.1: ``  `On shell', that is, assuming
the matter fields to satisfy their equations of motion, the matter field action is locally Weyl invariant if and only if the
corresponding energy-momentum tensor is traceless.''

We have stated the previous results in terms of classical Lagrangian densities. But as noted in the text, if these are generalized to trace dynamics
Lagrangian densities by making the fields matrix valued, adding an overall trace over the underlying Hilbert space, and symmetrizing Yukawa coupling
terms where needed, all of the manipulations described above go through for a classical metric $g_{\mu \nu}$.  The only change will be
that $T_0^0$ becomes the trace Hamiltonian density ${\bf H}={\rm Tr} T_0^0$, and so the trace Hamiltonian given by Eq. \eqref{eq:traceham} is
Weyl scale invariant.

The spinor  number current $N^{\mu}=\bar \psi \gamma^{\mu} \psi$, which  obeys $\nabla_{\mu} N^{\mu}=0$ on shell,
clearly obeys $\delta_{\omega} N^{\mu} = -4 \omega N^{\mu}$ off-shell, so the conserved fermion number $N=\int d^3x  (^{(4)}g)^{1/2} N^0$ is Weyl scale invariant off-shell.
These statements immediately carry over to the trace dynamics generalization ${\bf N}$.  We digress to remark that the corresponding scalar quantity
$M=\bar \psi \psi$ obeys $\delta_{\omega}M = -3 \omega  M$, and so the mass-like action term formed from this, $\int d^4x (^{(4)}g)^{1/2} M$
is not Weyl scale invariant.  Hence we expect that when spinor source terms are introduced, the induced effective action for the spinor sources will not acquire a mass term, in direct analogy with the exclusion of a true cosmological constant  term in the gravitational effective action. We expect this to play an important
role in the application of trace dynamics to building models unifying the standard model of particle physics with gravitation, since it will extend
the class  of models in which the generation of Planck scale masses is forbidden.

Returning to the conserved quantities appearing in the trace dynamics canonical ensemble, we consider finally the current $\tilde C^{\mu}$ associated
with the conserved operator $\tilde C$.  These \cite{adlerkempf} have the interpretation as the on-shell conserved current and charge associated
with global $U(N)$ invariance of the trace dynamics action.  The current $\tilde C^{\mu}$ is easily calculated by replacing all matrix fields
$q$ by the commutator $[\Lambda,q]$ and isolating the term ${\rm Tr}\partial_{\mu} \Lambda C^{\mu}$.  Applying this
recipe to the trace dynamics generalizations of the actions given above, we find the following.

(1)~~For the scalar field, we have
\begin{align}
\tilde C^{\mu}=& g^{\mu \alpha} [\phi, \partial_{\alpha} \phi]~~~,\cr
\tilde C^0= &g^{0 \alpha} [\phi, \partial_{\alpha} \phi] = [\phi,p_{\phi}]~~~.\cr
\end{align}

(2)~~For the Yang-Mills field, we have
\begin{align}
\tilde C^{\mu}=& -[A_{\lambda},F^{\mu \lambda}]~~~,\cr
\tilde C^0= & -[A_{\lambda}, F^{0 \lambda}]=[A_{\lambda}, p_{A_{\lambda}}]~~~.\cr
\end{align}

(3)~~For the Dirac spinor field, we have

\begin{align}
\tilde C^{\mu}=&-i\{ \bar \psi\gamma^{\mu}, \psi \}~~~,\cr
\tilde C^0= &-i\{ \bar \psi \gamma^0, \psi  \} =\{\psi, p_{\psi} \}~~~.\cr
\end{align}

In all three cases we see that off-shell $\delta_{\omega}\tilde C^{\mu} = -4 \omega \tilde C^{\mu}$, and so
\begin{equation}
 \delta_{\omega} [(^{(4)}g)^{1/2} \tilde C^{\mu}]=0~~~,
 \end{equation}
 which implies the Weyl scaling invariance of the conserved charge $\int d^3x (^{(4)}g)^{1/2} \tilde C^0$ appearing in the
 canonical ensemble.
\section{Construction and properties of the mixed index gravitational pseudotensor}

We show here how to construct a mixed index gravitational pseudotensor with the following properties.

\begin{enumerate}

\item When the metric is written as $g_{\mu \nu}=\eta_{\mu \nu} +h_{\mu \nu}$, with $h_{\mu \nu}$ not necessarily small, the tensor
is quadratic or higher order in $h_{\mu \nu}$.

\item The pseudotensor obeys the conservation law $\partial_{\nu} [{\surd g} (T_{\mu}^{\nu}+t_{\mu}^{\nu})]=0$, where for purposes
of this Appendix we abbreviate $\surd g \equiv (^{(4)}g)^{1/2}$.

\item For an isolated system, the three space integral $\int d^3x \surd g (T_{\mu}^0+t_{\mu}^0)$ gives the usual four-momentum $P_{\mu}$
defined by the asymptotic solution.

\item The matter component of the total energy, and the gravitational component of the total energy $\int d^3 x \surd g t_0^0$, are both
invariant under three space coordinate transformations $t \to t$, $\vec x \to \vec x(\vec x^{\,\prime})$.

\item The total gravitational energy for an isolated system is equal to that calculated from any pseudotensor obeying properites (1) and (2), such as the  Einstein-Dirac pseudotensor
$t_{\mu~{\rm ED}}^{\nu}$, that is, $\int d^3x \surd g t_0^0=\int d^3x \surd g t_{0~{\rm ED}}^0$.

\item For {\it linear} coordinate transformations, the Einstein-Dirac pseudotensor transforms as a tensor.

\item For the special case of a spatially flat universe using Cartesian coordinates, the Einstein-Dirac pseudotensor is spatially uniform,
taking the form of Eq. \eqref{eq:spatflat}, and  has a local physical significance.

\end{enumerate}

To prove these statements, we follow a constructive procedure given by Weinberg \cite{wein}, with modifications appropriate to the mixed index case and
 to include  factors
of $\surd g$ where needed.  We start from the mixed index form of the Einstein equations,
\begin{equation}\label{eq:einmixed}
G_{\mu}^{\nu}=-8\pi G T_{\mu}^{\nu}~~~,
\end{equation}
and separate $G_{\mu}^{\nu}$ into a part $G_{\mu}^{(1)\,\nu}$ and a remainder $\Delta G_{\mu}^{\nu}$,
\begin{equation}
G_{\mu}^{\nu}=G_{\mu}^{(1)\,\nu}+\Delta G_{\mu}^{\nu}~~~,
\end{equation}
so that $\Delta G$ is quadratic (and higher) order in $h_{\mu \nu}$.  Adopting the convention
that indices on first order quantities like $h_{\mu \nu}$, $\partial_{\mu}$ and $G_{\mu}^{(1)\,\nu}$
are raised and lowered with $\eta_{\mu \nu}$, we have explicitly
\begin{align}
G_{\mu}^{(1)\,\nu}=&\eta^{\nu \kappa} G^{(1)}_{\mu  \kappa}~~~,\cr
G^{(1)}_{\mu \kappa}=&R^{(1)}_{\mu \kappa}-\frac{1}{2}\eta_{\mu \kappa} R^{(1)\,\lambda}_{\lambda}~~~,\cr
\end{align}
and with \big(see Eq. (7.6.2) of \cite{wein}\big)
\begin{equation}
R^{(1)}_{\mu \kappa}=\frac{1}{2}\left(\frac {\partial^2 h^{\lambda}_{\lambda} }{\partial x^{\mu} \partial x^{\kappa}}
-\frac {\partial^2 h^{\lambda}_{\mu} }{\partial x^{\lambda} \partial x^{\kappa}}
-\frac {\partial^2 h^{\lambda}_{\kappa} }{\partial x^{\lambda} \partial x^{\mu}}
+\frac {\partial^2 h_{\mu \kappa} }{\partial x^{\lambda} \partial x_{\lambda}}
\right)~~~.
\end{equation}
We make a similar splitting for $\surd g$, by writing
\begin{equation}
\surd g = 1 + \Delta_{\surd g}~~~,
\end{equation}
so that $\Delta_{\surd g}$ is at least linear in $h_{\mu \nu}$.

Multiplying Eq. \eqref{eq:einmixed} by $\surd g$ and doing some algebraic rearrangement,
it can be rewritten in the form
\begin{equation}\label{eq:einmixed1}
G_{\mu}^{(1)\,\nu}=-8\pi G \surd g (T_{\mu}^{\nu} + t_{\mu}^{\nu})~~~,
\end{equation}
with
\begin{equation}
t_{\mu}^{\nu}=\frac{1}{8\pi G}\left[\Delta G_{\mu}^{\nu}+ \frac{\Delta_{\surd g}}{\surd g} G_{\mu}^{(1)\,\nu}\right]~~~.
\end{equation}
By construction, $t_{\mu}^{\nu}$ is at least quadratic in $h_{\mu}^{\nu}$,
and since $G_{\mu}^{(1)\,\nu}$ obeys the linearized Bianchi identity
\begin{equation}\label{eq:contractbian}
\partial_{\nu} G_{\mu}^{(1)\,\nu}=0~~~,
\end{equation}
we have
\begin{equation}
\partial_{\nu} [\surd g (T_{\mu}^{\nu}+ t_{\mu}^{\nu})]=0~~~.
\end{equation}
This construction completes the demonstration of properties (1) and (2) listed above.
We note, however, that we cannot rewrite $ t_{\mu}^{\nu}$ as a symmetric tensor by lowering the
index $\nu$ with $\eta_{\nu \kappa}$, because while this turns $ G_{\mu}^{(1)\,\nu}$ into
a symmetric tensor $G^{(1)}_{\mu \kappa}$, the quantity $\Delta G_{\mu}^{\nu}$ is the difference of a tensor $G_{\mu}^{\nu}$ that needs
the full metric $g_{\nu \kappa}$to lower the index $\nu$ to give a symmetric tensor, and of $G_{\mu}^{(1)\,\nu}$.

Following the discussion in \cite{wein}, we now use the fact that
$G_{\mu}^{(1)\,\nu}=\eta_{\mu \lambda} G^{(1)\,\nu \lambda}$ can be written as a total divergence,
\begin{equation}
G_{\mu}^{(1)\,\nu}=\partial_{\rho} (\eta_{\mu \lambda} Q^{\rho \nu \lambda})~~~,
\end{equation}
with $Q^{\rho \nu \lambda}$ given by Eq. (7.6.19) of \cite{wein},
\begin{equation}
Q^{\rho \nu \lambda}=\frac{1}{2}\left(\frac{\partial h_{\mu}^{\mu}}{\partial x_{\nu}}\eta^{\rho \lambda}
-\frac{\partial h^{\mu \nu}}{\partial x^{\mu}}\eta^{\rho \lambda}
+\frac{\partial h^{\nu \lambda}}{\partial x_{\rho}}-(\nu \leftrightarrow \rho)\right)
~~~.
\end{equation}
Since $Q^{\rho\nu \lambda}$ is antisymmetric in $\nu$ and $\rho$, the contracted Bianchi identity of
Eq. \eqref{eq:contractbian} is automatically satisfied.  Let us now form the volume integral
\begin{equation}
\int d^3x \surd g (T_{\mu}^{\nu}+t_{\mu}^{\nu})=-\frac{1}{8\pi G} \int d^3x G_{\mu}^{(1)\,\nu}
=-\frac{1}{8\pi G} \int d^3x\, \eta_{\mu \lambda} \partial_{\rho} Q^{\rho \nu \lambda}~~~.
\end{equation}
Defining the total four momentum by $P_{\mu} = \eta_{\mu \lambda} P^{\lambda}$,
we have
\begin{align}\label{eq:spatint}
P^{\lambda}=&\eta^{\lambda \mu} \int d^3x \surd g (T_{\mu}^{0}+t_{\mu}^{0})
=  -\frac{1}{8\pi G} \int d^3x  \partial_{\rho} Q^{\rho 0 \lambda}\cr
=& -\frac{1}{8\pi G} \int d^3x  \partial_i Q^{i 0 \lambda}
=-\frac{1}{8\pi G} \int dS_i Q^{i 0 \lambda}~~~,\cr
\end{align}
with the surface integral on the second line evaluated over the sphere at spatial infinity. This demonstrates property (3) listed above.
Note that once we have have identified $-\frac{1}{8\pi G}  \partial_i Q^{i 0 \lambda}$ as an expression of energy-momentum density, we can
similarly define a total angular momentum by
\begin{align}
J^{\nu \lambda}=   -\frac{1}{8\pi G} \int d^3x  \Big( x^{\nu} \partial_i Q^{i0\lambda}-x^{\lambda} \partial_i Q^{i0\nu}\Big)~~~,
\end{align}
and convert it to a surface integral over the sphere at infinity.  But because $t_{\mu}^{\nu}$ is not symmetric in its indices,
the integrand in this equation cannot be rewritten in terms of a locally conserved angular momentum four vector current density constructed
from $T_{\mu}^{\nu}+ t_{\mu}^{\nu}$.

Property (4) is a consequence of the facts that $\int d^3 x \surd g T_0^0$ is invariant under three space coordinate transformations
that keep the time $t$ fixed, since $d^3 x \surd g$ and $T_0^0$ both are invariant under these transformations, and that for an isolated
system with an asymptotically flat metric, the total energy
$P^0$ defined by the spatial integral in Eq. \eqref{eq:spatint} is also invariant under spatial coordinate transformations in the
interior (non-asymptotic) region.
Hence $\int d^3x \surd g t_0^0 =P^0- \int d^3 x \surd g T_0^0$ is invariant under such spatial
coordinate transformations. This was Dirac's \cite{dirac} motivation for including the $\surd g$ factor in his definition of
the gravitational stress-energy tensor.

The construction we have given for $t_{\mu}^{\nu}$ is not unique.  Suppose there is another $\tilde t_{\mu}^{\nu}$ that is at least quadratic
in $h_{\mu \nu}$ and obeys $\partial_{\nu}[ \surd g (T_{\mu}^{\nu}+\tilde t_{\mu}^{\nu})]=0$.  Forming the difference $\Delta t_{\mu}^{\nu} = \tilde t_{\mu}^{\nu} -
 t_{\mu}^{\nu}$, we have $\partial_{\nu} [\surd g \Delta t_{\mu}^{\nu}]=0$, which implies that
 \begin{equation}
 \surd g \Delta t_{\mu}^{\nu}= \partial_{\rho} D_{\mu}^{\nu \rho}~~~,
 \end{equation}
 with $D_{\mu}^{\nu \rho}$ antisymmetric in $\nu$ and $\rho$.  Then calculating the corresponding total gravitational energy difference, we have
 \begin{align}
 &\Delta P_{\mu} \propto \int d^3x \surd g \Delta t_{\mu}^0=  \int d^3x \partial_{\rho} D_{\mu}^{0 \rho} \cr
 =&\int d^3x \partial_i D_{\mu}^{0i} = \int dS_i D_{\mu}^{0i}=0~~~,\cr
 \end{align}
 since the fact that  $\Delta t_{\mu}^{\nu}$ is of quadratic or higher order in $h_{\mu}^{\nu}$ implies that for an isolated system,
 the surface integral at spatial infinity vanishes.  Hence one obtains the same {\it total} gravitational energy momentum
 from either $ t_{\mu}^{\nu}$ or $\tilde t_{\mu}^{\nu}$, even though they define different local energy-momentum
 distributions.  A particular elegant choice of $\tilde t_{\mu}^{\nu}$ has been given by Einstein and Dirac \cite{dirac}, and so we have demonstrated property (5) stated above. That is, the
 Einstein-Dirac pseudotensor $t_{\mu~{\rm ED}}^{\nu}$  given by
\begin{equation}
t_{\mu~{\rm ED}}^{\nu}=\frac{1}{16 \pi G (^{(4)}g)^{1/2} } [ (g^{\alpha \beta} (^{(4)}g)^{1/2} )_{,{\mu}} (\Gamma^{\nu}_{\alpha\beta}
-\delta^{\nu}_{\beta} \Gamma^{\sigma}_{\alpha \sigma}\big) - \delta^{\nu}_{\mu} g^{\alpha \beta} (^{(4)}g)^{1/2} \big( \Gamma^{\sigma}_{\alpha \beta}
\Gamma^{\rho}_{\sigma \rho} -  \Gamma^{\sigma}_{\alpha \sigma}  \Gamma^{\sigma}_{ \beta \rho}\big)] ~~~,
\end{equation}
 yields the same total $P_{\mu}$ for an isolated system as the $t_{\mu}^{\nu}$ constructed above, which we have shown gives the usual
 asymptotically defined energy-momentum for an isolated system.  Subtracting the matter energy-momentum, it also yields the same total gravitational contribution to $P_{\mu}$ as any pseudotensor obeying properties (1) and (2).

 Since the Einstein-Dirac pseudotensor is constructed in terms of the affine connection, it transforms as a tensor when the affine
 connection transforms as a tensor.  Because the inhomogeneous terms in the transformation of the affine connection under a coordinate
 transformation arise from second derivatives of the coordinate transformation, in the special case of linear coordinate transformations, the
 Einstein-Dirac pseudotensor transforms as a tensor.  This is property (6).

 Let us now consider a spatially flat Robertson-Walker universe ($k=0$), for which the line element in Cartesian coordinates is simply
 \begin{equation}
 ds^2=dt^2-a(t)^2(dx^2+dy^2+dz^2)~~~,
 \end{equation}
 corresponding to the metric components
\begin{equation}\label{eq:flat}
g_{00}=1~,~~g_{0i}=g_{i0}=0,~~g_{ij}=-a(t)^2 \delta_{ij}~~~.
\end{equation}
 An easy calculation shows that the only nonvanishing affine connection components are
 \begin{equation}
 \Gamma^{i}_{0j}=\frac{\dot a}{a} \delta^{i}_{j}~,~~ \Gamma^{0}_{ij}= a \dot a\, \delta_{ij}~~~.
 \end{equation}
 A further easy calculation then shows that the Einstein-Dirac pseudotensor takes the spatially
 uniform value
 \begin{align}
 t_{0~{\rm ED}}^0=&-\frac{3}{8 \pi G} \left( \frac{\dot a}{a}\right)^2~~~,\cr
 t_{0~{\rm ED}}^j=&0~~~,
 \end{align}
 which is property (7).  The spatial uniformity of this result could have been anticipated from property (6),
 since the isometries of the metric of Eq. \eqref{eq:flat}, which are spatial translations
 and rigid spatial rotations, are realized as linear coordinate transformations.
 Since the Friedmann equations tell us that
 \begin{equation}
 \left( \frac{\dot a}{a}\right)^2+\frac{k}{a^2}=\frac{8 \pi G}{3} \rho_{\rm tot}~~~,
 \end{equation}
 with $\rho_{\rm tot}$ the total matter contribution to the energy density,
 we see that for $k=0$ the sum of the gravitational and matter energy densities is zero.
( A related, more complicated calculation has been given by Mitra \cite{mitra} starting
from the Einstein form of the pseudotensor.  He concludes that a spatially flat universe
has zero total energy when it is static, but we do not find this restriction.  We have
not analyzed the reason for the discrepancy between his result and ours.)

For further discussion and properties of the mixed index pseudotensor, see \cite{goldberg} and
\cite{bergmann}.


\begin{thebibliography}{99}

\bibitem{adlermillard} S. L. Adler and A. C. Millard, Nucl. Phys. B{\bf473}, 199 (1996); other papers are cited in Ref. [3].
\bibitem{adlerkempf} S. L. Adler and A. Kempf, J. Math. Phys. {\bf 39}, 5083 (1998).
\bibitem{adlerbook}  S. L. Adler, {\it Quantum Theory as an Emergent Phenomenon: The Statistical Mechanics of Matrix Models
as the Precursor of Quantum Field Theory}, Cambridge University Press, Cambridge (2004).
\bibitem{adleressay2}S. L. Adler, ``Shadow Dark Matter as a Manifestation of $i \leftrightarrow -i$ Symmetry
in Pre-Quantum Trace Dynamics'', Honorable Mention in the 2013 Gravitation Essay Competition; submitted to International Journal of Modern Physics D.
\bibitem{adlersusy}  S. L. Adler, Nucl. Phys. B{\bf 499}, 569 (1997); S. L. Adler, Phys. Lett. B{\bf 407}, 229 (1997).
\bibitem{dyson} F. J. Dyson,   ``Is a Graviton Detectable?'', Poincar\'e Prize Lecture (2012), in press.
\bibitem{page} D. N. Page and C. D. Geilker, Phys. Rev. Lett. {\bf 47}, 979 (1981).
\bibitem{eppley} K. Eppley and E. Hannah, Found. Phys. {\bf 7}, 51 (1977).
\bibitem{dirac}  P. A. M. Dirac, {\it General Theory of Relativity}, Princeton University Press (1996), Secs. 31 and 32.
\bibitem{landau} L. Landau and E. Lifshitz, {\it The Classical Theory of Fields}, Addison-Wesley (1951), Sec. 11-9.
\bibitem{forger} M. Forger and H. R\"omer, Ann. Phys. {\bf 309}, 306 (2004).
\bibitem{adleressay1} S. L. Adler, Gen. Rel. and Grav. {\bf 29}, 1357 (1997).
\bibitem{muk} V. Mukhanov, {\it Physical Foundations of Cosmology}, Cambridge University Press (2005), Sec. 1.2.
\bibitem{parker} L. E. Parker and D. J. Toms, {\it Quantum Field Theory in Curved Spacetime}, Cambridge University Press, Cambridge (2009).
\bibitem{wein} S. Weinberg, {\it Gravitation and Cosmology: Principles and Applications of the General Theory of Relativity}, John Wiley \& Sons (1972),
pp. 165-171.
\bibitem{mitra} S. Mitra, Gen. Rel. Grav. {\bf 42}, 443 (2010).
\bibitem{goldberg} J. N. Goldberg, Phys. Rev. {\bf 111}, 315 (1958).
\bibitem{bergmann} P. G. Bergmann, Phys. Rev. {\bf 112}, 287 (1958).


\end{thebibliography}
\end{document}